\documentclass[namedreferences]{solarphysics}

\usepackage[hyperref,optionalrh]{spr-sola-addons} 
\usepackage{graphicx}        
\usepackage{amssymb}        
\usepackage[usenames,dvipsnames,svgnames]{xcolor}
\usepackage{url}             
\usepackage{rotating}

\newcommand{\angstrom}{\mbox{\normalfont\AA}}

\newcommand{\degree}{^{\circ}}
\newcommand{\eg}{{\it e.g.}}
\newcommand{\etal}{{\it et al.}}
\newcommand{\ie}{{\it i.e.}}

\newcommand{\BE}{\begin{equation}}
\newcommand{\EE}{\end{equation}}
\newcommand{\BA}{\begin{eqnarray}}
\newcommand{\EA}{\end{eqnarray}}
 \newcommand{\fig}[1]{Figure~\ref{#1}}

\renewcommand{\vec}[1]{\mathbf{#1}} 
\let\oldhat\hat  
\renewcommand{\hat}[1]{\oldhat{\mathbf{#1}}}  

\newcommand{\so}{\sigma_{\rm obs}}
\newcommand{\st}{\sigma_{t}}
\newcommand{\Vp}{V_{\parallel}}
\newcommand{\Vph}{V_{\varphi}}
\newcommand{\Vpr}{V'_{ r}}

\newcommand{\Vr}{V_{r}}
\newcommand{\Vt}{V_{\theta}}
\newcommand{\Wo}{W_{0}}

\newcommand{\Wph}{W_{\varphi}}
\newcommand{\Wpr}{W'_{r}}

\newcommand{\Wr}{W_{r}}

\newcommand{\myv}[1]{\hspace{-1pt}\textbf{\textit{#1}}}

\chardef\us=`\_

\begin{document}

\begin{article}
\begin{opening}

\title{Apparent and Intrinsic Evolution of Active Region Upflows}

\author[addressref={1},corref,email={deborah.baker@ucl.ac.uk}]{\inits{D.}\fnm{Deborah}~\lnm{Baker}\orcid{0000-0002-0665-2355}}
\author[addressref={2},email={mjanvier@ias.u-psud.fr}]{\inits{M.}\fnm{Miho}~\lnm{Janvier}\orcid{0000-0002-6203-5239}}
\author[addressref={3},email={Pascal.Demoulin@obspm.fr}]{\inits{P.}\fnm{Pascal}~\lnm{D\'emoulin}\orcid{0000-0001-8215-6532}}
\author[addressref={4,5},email={mandrini@iafe.uba.ar}]{\inits{C.H.}\fnm{Cristina H.}~\lnm{Mandrini}\orcid{0000-0001-9311-678X}}

\runningauthor{D.~Baker \etal}
\runningtitle{Active Region Upflows}

\address[id={1}]{University College London - Mullard Space Science Laboratory, Holmbury, St. Mary, Dorking, Surrey, KT22 9XF, U.K.}
\address[id={2}]{Institut d'Astrophysique Spatiale, CNRS, Univ. Paris-Sud, Universit\'e Paris-Saclay, B\^at. 121, 91405 Orsay cedex, France}
\address[id={3}]{Observatoire de Paris, LESIA, UMR 8109 (CNRS), F-92195 Meudon Principal Cedex, France}
\address[id={4}]{Instituto de Astronom\'\i a y F\'\i sica del Espacio (IAFE), UBA-CONICET, Buenos Aires, Argentina}
\address[id={5}]{Facultad de Ciencias Exactas y Naturales (FCEN), UBA, Buenos Aires, Argentina}

\begin{abstract}
We analyze the evolution of Fe {\sc xii} coronal plasma upflows from the edges of ten active regions (ARs) as they cross the solar disk using the \emph{Hinode Extreme Ultraviolet Imaging Spectrometer} (EIS).  
Confirming the results of D\'emoulin \etal~(2013, {\it Sol. Phys.} {\bf 283}, 341), we find that for each AR there is an observed long term evolution of the upflows which is largely due to the solar rotation progressively changing the viewpoint of dominantly stationary upflows.  
From this projection effect, we estimate the unprojected upflow velocity and its inclination to the local vertical.  
AR upflows typically fan away from the AR core by 40$\degree$ to near vertical for the following polarity.  The span of inclination angles is more spread for the leading polarity with flows angled from -29$\degree$ (inclined towards the AR center) to 28$\degree$ (directed away from the AR). 
In addition to the limb--to--limb apparent evolution, we identify an intrinsic evolution of the upflows due to coronal activity which is AR dependent. Further, line widths are correlated with Doppler velocities only for the few ARs having the largest velocities.  
We conclude that for the line widths to be affected by the solar rotation, the spatial gradient of the upflow velocities must be large enough such that the line broadening exceeds the thermal line width of Fe {\sc xii}. 
Finally, we find that upflows occurring in pairs or multiple pairs is a common feature of ARs observed by \emph{Hinode}/EIS, with up to four pairs present in AR 11575.
This is important for constraining the upflow driving mechanism as it implies that the mechanism is not a local one occurring over a single polarity.  
AR upflows originating from reconnection along quasi-separatrix layers (QSLs) between over-pressure AR loops and neighboring under-pressure loops is consistent with upflows occurring in pairs, unlike other proposed mechanisms acting locally in one polarity.
\end{abstract}
\end{opening}

\section{Introduction}
    \label{intro}
\subsection{General Characteristics of Active Region Upflows}
Since the launch of  the \emph{Hinode Extreme Ultraviolet Imaging Spectrometer} (EIS), steady coronal plasma upflows have been observed emanating from the edges of active regions \citep[ARs;][] {sakao07,harra08}.
Their blue-shifted velocities are in the range [5, 50] km s$^{-1}$ for the Fe {\sc xii} 195~\angstrom~emission line.
Similar velocities have been measured at much higher spatial resolution by the {\it Interface Region Imaging Spectrograph} (IRIS) also using Fe {\sc xii} \citep{testa16}.
Some ARs show large blue wing asymmetries exceeding 100 km s$^{-1}$ \citep{bryans10,depontieu10,tian11,brooks12,vanninathan15}.
Upflow temperatures range from 1 -- 2.5 MK \citep{warren11} and the Doppler velocities increase with temperature in the corona  \citep{delzanna08}. 

Typically, these large-scale upflows are located over monopolar regions of strong magnetic field \citep{doschek08,delzanna08}. 
Upflowing plasma is observed by \emph{Hinode}/EIS in approximately the same locations for at least 8--9 days as ARs transit the solar disk \citep{demoulin13}.
Recently, \cite{zangrilli16} have shown that the upflows persist for much longer periods.
Using data from {\it Solar and Heliospheric Observatory's Ultraviolet Coronagraph Spectrometer} (SOHO/UVCS), they  demonstrated that upflows from AR 8100 extended into the intermediate corona to become outflows and persisted for the AR's entire lifetime which spanned five solar rotations.

\subsection{Upflow Driving Mechanisms}
\label{mechanisms}
\citet{baker09}, \citet{lvdg12}, \citet{demoulin13}, and \citet{mandrini15} have demonstrated that upflows occur along quasi-separatrix layers (QSLs) which are thin 3D volumes where magnetic field lines display strong gradients in magnetic connectivity \citep{demoulin96}.  
QSLs are preferential locations for current sheet development and magnetic reconnection (\eg\ \citealp{aulanier06}; see \citealp{janvier17} for a review).  QSLs are typically present between the core of the AR and surrounding regions (including open field). The AR core has over-pressure loops compared to the loops in these nearby regions. As the AR is developing, its core grows so that associated magnetic field motions can build current layers at the QSLs. This process is expected to drive a long-term reconnection process (ranging from quasi-continuous to a series of frequent small reconnection events). After reconnection, the resulting pressure gradient drives the upflows along the reconnected field lines.
 
A similar mechanism was proposed by \citet{delzanna11} whereby upflows were related to a coronal null-point of a pseudostreamer.
Interchange reconnection ({\it i.e.} taking place between AR loops and open field lines) can take place at the coronal null, creating a steep pressure gradient and a rarefaction wave in the reconnected loops \citep{bradshaw11}.
As QSLs are the generalized form of separatrices and null points in 3D, and include them, reconnection at these topological locations is a subset of QSL reconnection.
Clearly, magnetic reconnection is likely to play a key role in driving AR plasma upflows.

Several other mechanisms have been proposed as direct or indirect drivers of AR upflows observed by \emph{Hinode}/EIS such as AR expansion \citep{murray10}, waves \citep[\eg\ ][]{wang09,verwichte10,ofman12,galsgaard15}, coronal plasma circulation \citep{marsch08}, chromospheric evaporation \citep{delzanna08}, and type II spicules \citep[\eg\ ][]{depontieu09,depontieu10,tian12}, among others.  

\subsection{Paper Road Map}

\cite{demoulin13} used \emph{Hinode}/EIS data to derive the physical properties of the large-scale upflows observed on both sides of AR 10978 as it crossed the solar disk.  
A least-squares fitting of a stationary flow model to the data provided the means to distinguish apparent evolution due to line-of-sight projection effects from intrinsic flow evolution.
It was shown that the upflows have a strong collimated stationary component.

In this study we apply the same analysis as in \cite{demoulin13} to nine additional ARs with good limb--to--limb EIS coverage.  
Section \ref{paper1} provides a pr\'ecis of the steady-flow model and key results from AR 10978.  
In Section \ref{obs} we summarize the EIS observations in our AR sample.
We analyze the limb--to--limb Fe {\sc xii} upflow evolution for three case studies in Section \ref{model_results}.  
Then we generalize the results for all ARs before turning to the evolution of line widths in Section \ref{ntvel}.  
We compare the inclinations of the stationary flows of AR 10926 with those of the coronal magnetic field using a linear force-free field extrapolation of the AR photospheric line-of-sight (LOS) field (Section \ref{extrap}). 
   In Section \ref{pairs}, we show that upflows in AR 11575 occur in four pairs which lets us investigate any similarity between the two polarities of the same pair.
Finally, we summarize our results and draw our conclusions (Section \ref{end}).

\section{Steady-Flow Model Applied to AR 10978}
\label{paper1}
\subsection{Steady-Flow Model}
\label{P1_model}
\cite{demoulin13} analyzed the limb-to-limb evolution of plasma upflows from AR 10978 using \emph{Hinode}/EIS.  
A global evolution of the AR's upflow Doppler velocities is apparent as it crosses the solar disk from 6 -- 16 December 2007. 
On the AR's eastern side, upflow velocities increase almost continuously with time before reversing early on 15 December.
A similar evolution takes place on the western side but the change from increasing to decreasing upflow velocities occurs four days earlier. 
Such a coherent, large-scale evolution is clear evidence of a LOS projection effect evolving with the AR's position on the disk, thereby demonstrating that the upflows have a strongly collimated stationary component. 
Upflow velocities on either side of the AR peak when the stationary component is parallel to the LOS of \emph{Hinode}/EIS.

A stationary flow model was developed to quantify the effect of the evolving LOS projection in order to separate apparent velocity evolution, due to solar rotation, from intrinsic velocity evolution, due to some form of activity such as flux emergence, flares, and coronal mass ejections (CMEs).  
In applying this simple model to EIS Doppler-velocity data of AR 10978, it was assumed that the plasma upflows are continuously driven with the same velocity magnitude and orientation in the local solar frame.  
The main equations of the model are presented here for convenience.  
See \cite{demoulin13} for a complete account of the model and its underlying assumptions, uncertainties, and limitations.  
  
In spherical coordinates with the radial direction [$r$], the longitude [$\varphi$], and the latitude [$\theta$], the velocity [\myv{V}] is:
\begin{equation}
\vec{\textit{\textbf{V}}} = V_{r}\hat{u}_{r} + V_{\varphi}\hat{u}_{\varphi} + V_{\theta}\hat{u}_{\theta},
\end{equation}
where $\hat{u}_{r}$, $\hat{u}_{\varphi}$, $\hat{u}_{\theta}$ are unit vectors.
The position of the flow on the solar disk, normalized to the solar radius, is given by $X = \sin \varphi \cos \theta$ in the east-west direction and $Y = \sin \theta$ in the north\,--\,south direction.
The observed velocity component, $\Vp$, is
\begin{eqnarray}
                 \Vp &=& \Vpr \sqrt{1 - X^2 - Y^2} 
                      - \Vph \frac{X}{1 - Y^2}       \label{equation2} \\
\mbox{with }\, \Vpr &=& \Vr - \Vt \frac{Y}{\sqrt{1 - Y^2}} \nonumber .
\end{eqnarray}

Intrinsic or temporal evolution of the stationary flows is indicated by deviations of the data from the fit to Equation (\ref{equation2}).

\subsection{Key Results - AR 10978}
\label{P1_results}
The inclination and angular spread of the strongest upflows in three coronal lines, Si {\sc vii}, Fe {\sc xii}, and Fe {\sc xv} (log T $\approx$ 5.8, 6.2, and 6.3, respectively), are determined using the flow model for AR 10978.  
The deduced 3D geometry of the flows is consistent with that of a thin, fan-like structure, in agreement with a magnetic field extrapolation of AR 10978.  
Fans on either side of the AR are tilted away from its core.  
A coherent fan structure is found on the eastern side (following polarity) of the AR where the strongest flows are stationary, unlike on the western side (leading polarity), where in addition to a stationary component, there is clear intrinsic evolution related to new flux emergence on the sides of this polarity \citep{mandrini15}. 

A similar dependence of line widths to that of Doppler velocities with respect to position of AR 10978 on the solar disk was previously found \citep{doschek08,bryans10}. 
The contribution to the  
line width, [$\Wo$], which is independent of the AR's position, is very small, implying that the line width is primarily due to a large dispersion of velocities in the  direction of the stationary flows 
(see Section \ref{ntvel} of this paper and Section 4.4 of \cite{demoulin13} for a full explanation of $\Wo$).
Taken in their entirety, these results indicate that the global picture is one where the same upflows are detected in a number of spectral lines at coronal temperatures. 
Blue-shifted plasma flows away from the AR's core along magnetic field lines within a narrow angular range but with a broad velocity range.  
The flows form a fan-like structure on either side of the AR as the collimated upflows expand with increasing height within the corona \citep[see Figure 14 of][]{demoulin13}.
The upflow driver(s) is(are) acting for extended periods, at least as long as the time it takes for the AR to cross the solar disk. The location of the flows within the magnetic configuration coupled with their narrow angular extent are compatible with a flow resulting from reconnection at QSLs, including separatrices \citep{baker09,delzanna11,lvdg12,demoulin13,mandrini15}.

\begin{table}[t]
\renewcommand{\arraystretch}{1.1}
  \caption{ARs included in the limb-to-limb upflow study.  The minimum age is given by when the ARs are first observed at the East limb or on the far-side. 
  }
\begin{tabular}{ccc}
  \hline
NOAA & Solar Object Locator & Min. Age \\
AR & (SOL) at CMP & (Days) \\
\hline
$10926$ &
2006-12-01T16:00:00L138C097&
8 
\\
$10938$ &
2007-01-18T16:00:00L226C086&
8 
\\
$10942$ &
2007-02-22T08:00:00L130C102&
8 
\\
$10953$ &
2007-05-01T17:00:00L308C100&
7 
\\
$10961$ &
2007-07-01T19:00:00L220C109&
7 
\\
$10978$ &
2007-12-11T22:00:00L226C102&
9 
\\
$11389$ &
2012-01-03T17:00:00L082C111&
21 
\\
$11564$ &
2012-09-05T21:00:00L071C104&
23 
\\
$11575$ &
2012-09-24T21:00:00L181C084&
22 
\\
$11589$ &
2012-10-15T17:00:00L266C079&
21 
\\
\hline
\end{tabular}
\label{info}
\end{table}

\begin{table}[t]
\renewcommand{\arraystretch}{1.1}
  \caption{Global properties of ARs included in the limb-to-limb upflow study. 
  }
\begin{tabular}{cccc}
  \hline
NOAA & Following & Leading  & Tot. Unsigned Flux\\
AR   & Polarity  & Polarity &  at CMP (Mx)\\
\hline
$10926$ &
Dispersed&
Strong spots $+$ dispersed&
$7.4\times 10^{21}$
\\
$10938$ &
Dispersed&
Dispersed&
$9.8\times 10^{21}$ 
\\
$10942$ &
Dispersed&
Dispersed&
$3.5\times 10^{21}$
\\
$10953$ &
Dispersed&
Strong spot&
$1.6\times 10^{22}$
\\
$10961$ &
Dispersed&
Strong spots&
$5.9\times 10^{21}$
\\
$10978$ &
Spots $+$ dispersed&
Spots $+$ dispersed&
$1.9\times 10^{22}$
\\
$11389$ &
Dispersed&
Strong spot $+$ dispersed&
$1.5\times 10^{22}$ 
\\
$11564$ &
Dispersed&
Dispersed&
$1.5\times 10^{22}$
\\
$11575$ &
Dispersed&
Strong spot&
$1.0\times 10^{22}$ 
\\
$11589$ &
Dispersed&
Dispersed&
$1.4\times 10^{22}$
\\
\hline
\end{tabular}
    \label{info2}
\end{table}

\section{Data and Analysis Method}
\label{obs}
\subsection{Observations}
\label{data}
Nine new ARs are considered in this study, along with the results previously found for AR 10978.
Observations are from the EIS instrument on board \emph{Hinode} \citep{culhane07}.  
ARs are selected from the EIS catalog based on how well their upflow regions are observed from limb to limb.  
In six cases, following polarities (FPs) and leading polarities (LPs) of the ARs are covered, but for four ARs only one polarity is within the EIS FOV throughout the disk transit time.

All of the ARs, with exception of AR 11575, have a relatively simple bipolar magnetic structure and flux distribution, which is consistent with that of ARs in the decay phase of their lifetimes. 
LPs range from highly concentrated spots to dispersed magnetic field regions, while FPs consist of only dispersed field with the exception of AR 10978 which contains coherent spots in both.
Nine of the ten ARs are classified as large, with a total unsigned flux greater than 5 $\times$ 10$^{21}$ Mx \citep{lvdg15}. 

NOAA number, Solar Object Locator (SOL) at the date and time of central meridian passage (CMP), and estimated minimum AR age are given in Table \ref{info}.
It is not possible to directly determine the age of any of the ARs since none were observed during flux emergence.
Instead, we provide a minimum age of the ARs solely based on when they are first detected at the Sun's East limb or on the far-side using observations of the {\it Solar Terrestrial Relations Observatory} (STEREO) A and B spacecraft.  Minimum ages range from 7 to 23 days.
The description of magnetic flux distribution for LPs and FPs, and total AR unsigned flux at CMP are given in Table \ref{info2}.
Information and results pertaining to AR 10978 from \cite{demoulin13} are included in the tables of this article for comparison purposes.

\subsection{Data Reduction}

The rastered images are constructed with the 1"~and 2"~slits in the scanning mode.
A total of 24 different EIS studies with varying exposure times, fields of view (FOV), and spectral line lists are used in this investigation.
See Table \ref{EIS_study} of the Appendix for EIS study details including study number, FOV, slit size, exposure time, and total raster time.
Data reduction is performed with  standard SolarSoft EIS routines which correct for dark current, hot, warm, and dusty pixels, detector bias, and cosmic rays. 
Further corrections are made for slit tilt and spectrum drift due to temperature variation throughout the spacecraft's orbit. 

Coronal spectral line profiles have been shown to have an extended blue wing in the upflow regions for a few ARs, \eg~AR 10978 in \citet{bryans10} and \citet{brooks12}, AR 10938 in \citet{hara08} and \citet{peter10}, both of which are included in our AR study, and AR 11123 in \citet{vanninathan15}. 
All of these authors required a second velocity component to fit the asymmetric Fe {\sc xii--xv} profiles as they found significant deviations from single Gaussian fits.
In most cases, a double Gaussian fit was sufficient to disentangle the primary and secondary velocity components of the AR upflows.
Typical primary component velocities were 5--20 km s$^{-1}$, whereas the secondary component velocities exceeded 100 km s$^{-1}$ in all cases.
Enhanced blue wing velocities have been linked to, among other phenomena, jets \citep[\eg][]{vanninathan15} and type II spicules \citep[\eg][]{mcintosh09,peter10}.
In \citet{demoulin13}, the results of double Gaussian fittings of the faster flows of AR 10978 could be trusted for a very limited number of pixels where upflow velocities were high enough to separate the secondary and primary components and where the emission measure was large enough to provide a reliable secondary component.  
For the other ARs in our study the variety of exposure times in the range 5\,--\,70 s proved to be problematic for consistently separating the components for all data sets. 

In addition to enhanced blue wings, the Fe {\sc xii} 195.12~\angstrom~line is blended with Fe {\sc xii} at 195.18~\angstrom. 
The self-blend must be taken into account in regions of high density $>10^{10}$ cm$^{-3}$, \eg~in AR cores \citep{young09}; however, in upflow regions densities are typically $\lesssim10^{9}$ cm$^{-3}$ \citep{bryans10,culhane14} so that we can ignore the effects of this blend.
Hence, in the study presented here, single Gaussian functions are fitted to the Fe {\sc xii} spectra in order to determine the total line intensity, the line width (full width at half maximum, FWHM), and the centroid wavelength. 

  \begin{figure}   
   \centerline{\includegraphics[width=0.6\textwidth,clip=0]{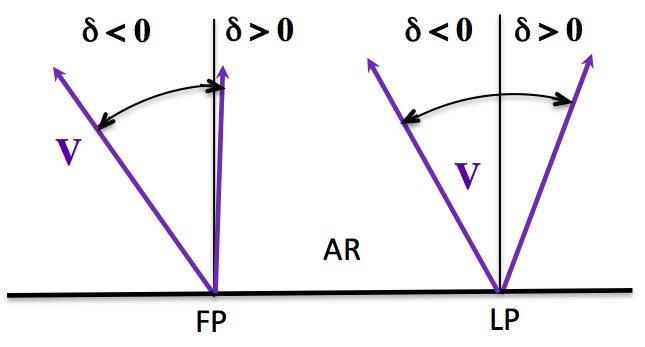}
              }
              \caption{Diagram to illustrate the sign and the typical range of the inclination angle {[$\delta$]} of the upflows with respect to the vertical for the leading (LP) and following polarities (FP) of an AR. The arrows show possible directions of the mean upflow velocity [\textit{\textbf{V}}] in each polarity.  }
   \label{inclination_fig}
   \end{figure}

\subsection{Deriving Steady Flows}
\label{flows}
We apply the steady-flow model summarized in Section \ref{P1_model} to upflow streams associated with the LPs and/or FPs of all ARs in our study. 
If the flows are stationary and well collimated, the observed velocity from a given location within the AR 
is described by the model (Equation (\ref{equation2})).
However, upflows have an unknown 3D geometry in the corona and therefore it is not possible to follow an upflow coming from a fixed AR location (within the local solar frame).
Instead, we proceed statistically over a full upflow region.  
If this region is defined by a given isocontour of $\Vp$, its extent is modified by the projection along the LOS (the selected region would be larger as the velocity is more along the LOS). 
As a collimated upflow region has a relatively small extension in latitude and longitude and therefore $\Vp$ is affected by a similar projection factor, the region(s) with the strongest flows is(are) essentially independent of the projection.
In practice, only a fixed number of pixels with the strongest upflows are retained.  
This does not fully disentangle the effect of the variable extension with the evolving projection but it is the most satisfactory method we have found \citep[see Section 2.4 of][]{demoulin13}.
We experimented with the number of data points from each raster observation in the range 20\,--\,100. 
We find that the effect on the derived mean velocities, line widths, and inclination angles is weak, which is consistent with the findings of \cite{demoulin13}. 

A classical convention is that $\Vp$ is positive for flows going away from the observer, so upflows have $\Vp<0$. 
For convenience, plots show -$\Vp$, \emph{i.e.} positive values. 
The highest 50 values are selected in each data set and the least-squares fitting to Equation (\ref{equation2}) is applied to these data points.  
Mean velocity [$V$], mean line width [$W$], and mean inclination angle [$\delta$] are determined from the fits for all available upflow regions. 
The angle $\delta$ measures the east-west inclination of the flows to the local vertical with the same sign convention in the FP and LP (\fig{inclination_fig}). 
The notations are the same as in \citet{demoulin13} where more explanations can be found.

\section{Steady-Flow Model Results}
\label{model_results}
\begin{figure}   
   \centerline{\includegraphics[width=1.0\textwidth,clip=0]{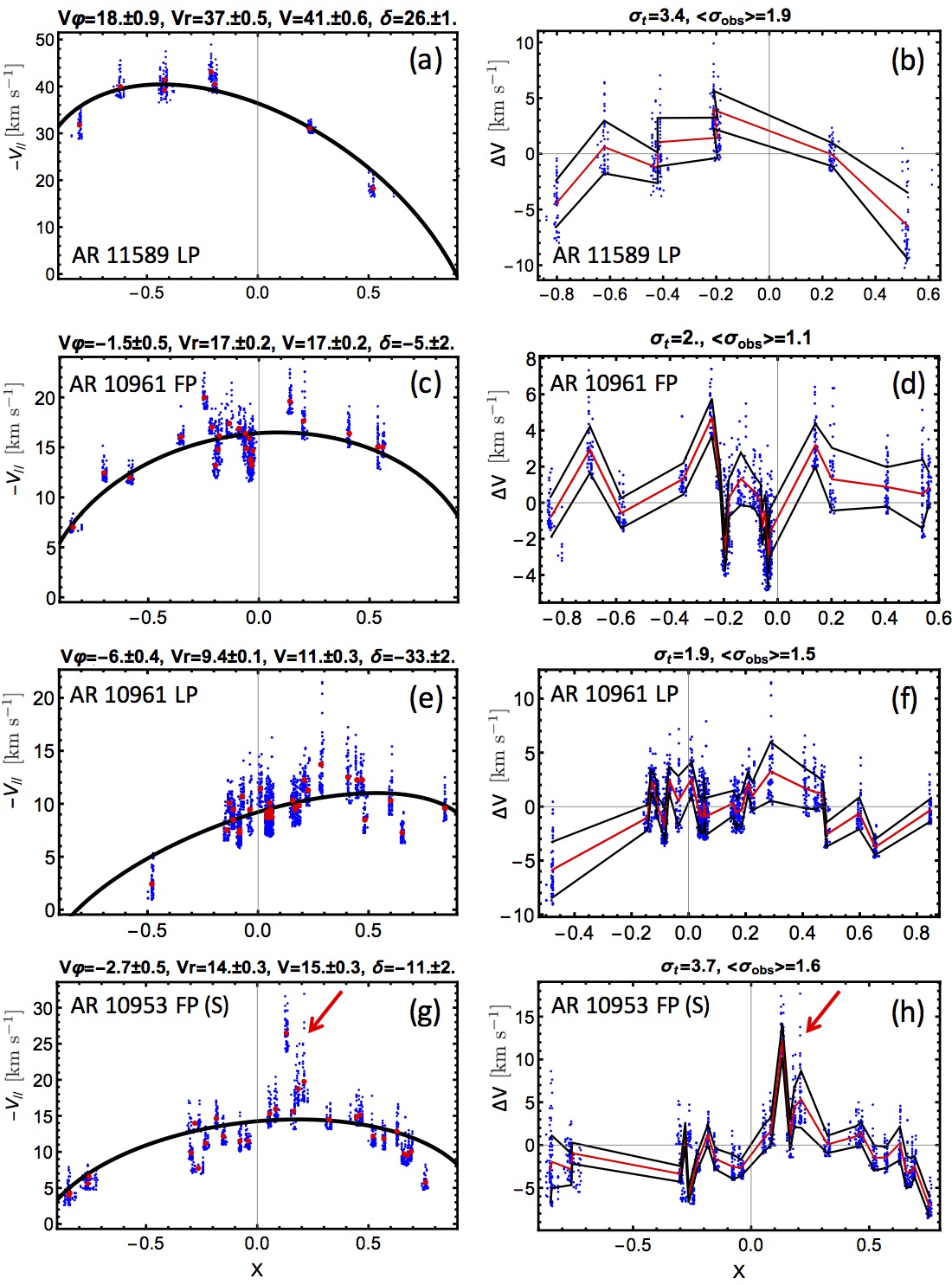}
              }
              \caption{Left panels (top to bottom): Steady-flow model results for ARs 11589, 10961 (FP and LP), and 10953 S (southern bipole) showing the dependence of the LOS upflow velocity [$\Vp$] from the eastern to the western AR position normalized to the solar radius [$X$].  Blue dots indicate the 50 strongest -$\Vp$ values selected in each data set and red dots are mean values for each data set. The black line is the least-squares fitting of Equation (\ref{equation2}) to these data points (in blue). Fit results are given at the top of each panel.  Right panels: Fluctuations of the velocities around the fitted values corrected for projection effects.  The red lines show the mean values of the fit difference and the black lines correspond to $\pm$ one standard deviation.  $\st$ is the standard deviation of all mean values (red) and $<\so >$ is the mean of the standard deviation for each data set. The red arrows point to the location of AR 10953 when the eruptions occurred (see Section~\ref{ar10953}).  $V$ and $\delta$ at the top of panel (g) are slightly different than in Table \ref{results_tab} because the fits include the activity times.  
              }
   \label{results_fig}
   \end{figure}

Examples of the fitting to the steady-flow model for selected ARs are displayed in Figure \ref{results_fig}.
In the left panels, -$\Vp$ {\it vs.} the $X$ position is plotted for four cases. 
Blue dots indicate the 50 highest -$\Vp$ values selected in each data set and red dots are their mean values.
Black lines represent the least-squares fittings of Equation (\ref{equation2}) to the data points.
Fit results are given at the top of each panel for these ARs and also provided for all ARs in Table \ref{results_tab}.
 

\begin{table*}[t]
\renewcommand{\arraystretch}{1.1}
 \caption{Results of the steady-flow model for ten ARs, including AR 10978. 
See \fig{inclination_fig} for the sign convention of $\delta$. The values indicate the results of the fits where the intrinsic activity has been removed, so as to only keep the apparent evolution parameters of the ARs.
NA in the different columns stands for data not available. Non-physical values are also found for the line broadening, meaning that the fit did not work in such cases.  See Section \ref{ntvel} for an explanation of $\Wo$, $\Wr$, and $\Wph$.  For AR 10978 FP, $\Wo$ is abnormally small so we only report ``large''.
   }
  \begin{tabular}{lcccccc}
  \hline
NOAA AR&
\multicolumn{2}{c}{Velocity (km s$^{-1})$}&
\multicolumn{2}{c}{($\Wr^2+\Wph^2)^{1/2}/\Wo$}&
\multicolumn{2}{c}{Inclination $\delta$ (deg.)}
\\
&FP&LP&FP&LP&FP&LP
\\
\hline
$10926$ &
18&
14&
0.29&
0.45&
-40&
-12
\\
$10938$ &
17&
NA&
0.84&
NA&
-4&
NA
\\
$10942$ &
14&
NA&
0.34&
NA&
4&
NA
\\
$10953$ (N) &
17&
19&
1&
0.55&
-8&
26
\\
$10953$ (S) &
14&
NA&
0.56&
NA&
-10&
NA
\\
$10961$ &
17&
11&
0.37&
1.39&
-5&
-29
\\
$10961$ &
NA&
9&
NA&
0.69&
NA&
21
\\
$10978$ &
32&
34&
large&
3.39&
-39&
19
 \\
$11389$ &
18&
NA&
1.53&
NA&
-10&
NA
\\
$11564$ &
31&
30&
0.55&
1.96&
-29&
28
\\
$11575$ (N) &
12&
12&
non-physical&
0.43&
-19&
-1
\\
$11575$ (C) &
12&
16&
0.18&
0.46&
-27&
9
\\
$11575$ (S) &
11&
20&
0.85&
2.29&
2&
11
\\
$11589$ &
NA&
41&
NA&
0.32&
NA&
26
\\
\hline
\end{tabular}
\label{results_tab}
\end{table*}

Fluctuations of the velocities around the fitted values are displayed graphically in the right panels.  
Data points (blue dots) have been corrected for projection effects using the fitted curves in the corresponding panels on the left.
Red lines represent the mean values of the fit differences and the black lines are $\pm$ one standard deviation.  
This gives a straightforward view on the fluctuation level and provides a direct way to identify intrinsic activity (see for example red arrow in \fig{results_fig}h).
At the top of the plots we indicate two measures of the  variation of the data values: the standard deviation [$\st$] of all mean values (shown in red in the panels) and the mean of the standard deviation [$<\so>$] for each data set.
They measure the fluctuations of the mean velocities with time and the mean value of the spatial fluctuations, respectively (for the largest velocities).

In the following subsections, we use the least-squares fitting of the stationary flow model to distinguish apparent from intrinsic flow evolution. Deviations of the data from the model indicate the temporal evolution of the flows.  
We give a brief account of the results for two ARs that have large and small values of the derived velocities, respectively, yet both display a dominant apparent evolution.
Then we look in more detail at an AR where an eruption occurred during its disk transit.

\subsection{Apparent Evolution - AR 11589 and AR 10961} 
\label{apparent} 
\emph{Hinode}/EIS observed the western upflow region of AR 11589 from 10\,--\,17 October 2012. 
Throughout this period, the dispersing magnetic flux of both polarities spreads over an ever-increasing area. The decaying AR has a bipolar magnetic configuration.  Loops originating in the upflow region over the positive polarity (FP) connect externally with the negative polarity of the nearby AR 11592 to the north\,--\,east and the surrounding quiet Sun. 
The upflow region on the negative polarity (LP) extends into the negative field of a coronal hole immediately to the west. 
Activity is steady with 23 C-class flares attributed to the AR; however, no significant internal flux emergence or CMEs are detected. 
 
The steady-flow model fit and the deviation from the fitting are shown in Figures~\ref{results_fig}a and \ref{results_fig}b, respectively.
From the curvature of the black line in the left plot, it is clear that the limb-to-limb evolution of upflows of the LP is dominated by the effects of the velocities projected onto the LOS.
The derived mean velocity is $41$ km s$^{-1}$, which is the strongest in our study.  
The error of this parameter, derived from the fitted parameters, is estimated to be $\pm 0.6$  km s$^{-1}$.
Overall, the model fit is very stable as the mean values of the fit differences (red line in Figure~\ref{results_fig}b) are $\lesssim 5$ km s$^{-1}$ until the AR approaches the western limb, where there are no data beyond approximately the +0.5 normalized AR position.
The temporal dispersion [$\st$] is 3.4 km s$^{-1}$ and the spatial mean dispersion [$<\so >$] is 1.9 km s$^{-1}$, so a factor 10 and 20 times smaller than the mean velocity.

Both polarities of AR 10961 were observed by \emph{Hinode}/EIS from 27 June to 6 July 2007.  For most of the  transit time, the isolated AR magnetic configuration is bipolar, though unlike AR 11589, its LP contains a strong, coherent spot.  Its total unsigned flux is $5.9 \times 10^{21}$ Mx.  
Activity is limited to a few jets and B-class flares in close proximity to the upflow region on the LP.  In Figure \ref{results_fig}, middle panels, the upflow streams on both sides of the AR exhibit a moderately strong dependence of $\Vp$ on the normalized X-position of the AR. 
The mean velocity for the FP is $17.0\pm 0.2$ km s$^{-1}$ and $11.0 \pm 0.3$ km s$^{-1}$ for the LP.  
The model fits are very good on either side of the AR.  
Values of $\st$ and $<\so>$ are similar for each polarity and much lower than those for AR 11589, while the values relative to the mean velocity are comparable.

In summary, AR 11589 and AR 10961 upflows exhibit a strongly collimated stationary component as they cross the solar disk so that the observed evolution is dominantly due to projection.  
This is the case for both low (AR 10961) and high (AR 11589) -$\Vp$ values.  In our data sample, this result is also present for the other ARs exhibiting low levels of activity. 

\subsection{Intrinsic Evolution - AR 10953 }\label{ar10953}
\begin{figure}   
\centerline{\includegraphics[width=0.7\textwidth,clip=0]{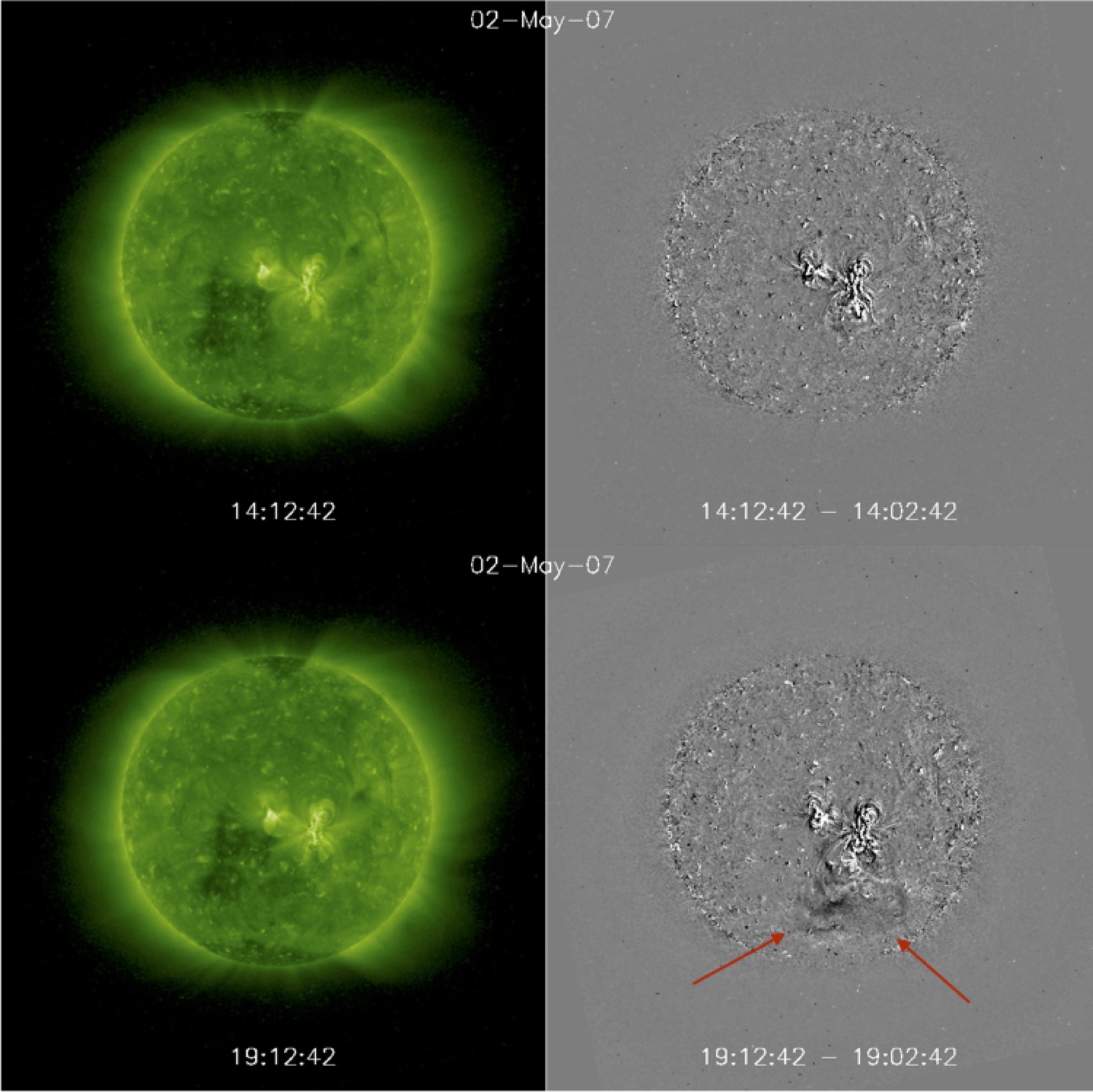}
             }
            \caption{STEREO B 195~\angstrom\ intensity images (left panels) and running difference images (right panels) of AR 10953 on 2 May 2007.  Top/bottom panels are before/during the first CME eruption.  Red arrows indicate the front edge of the CME in the running difference image at $\approx$19:00 UT.  The images are extracted from Movie 1. }
  \label{erupt}
\end{figure}

AR 10953 crossed the solar disk from 25 April to 6 May 2007.  
The AR had a simple bipolar configuration with a concentrated leading sunspot and dispersed FP.  
No significant flux emergence was observed.  
Flaring activity was limited to two C-class flares; however, there were several filament eruptions.
Two of them occurred at 18:30 UT and 23:12 UT on  2 May \citep{okamoto08,su09}.  
Figure \ref{erupt} shows STEREO B 195~\angstrom\ intensity and running difference images before and during the first eruption towards the southern edge of the AR.  
The images are taken from the included STEREO B movie, called Movie 1.
 
There was extensive \emph{Hinode}/EIS coverage of the upflow regions on both sides of the AR.  
On the east side, where the magnetic field is dispersed, there are two distinct upflow areas (AR 10953 (N) and (S) in Table \ref{results_tab}). 
The stationary flow model results for the FP southern region are displayed in the bottom panels of Figure \ref{results_fig}.  
Once again, there is an apparent flow evolution as is evident from the black line of the model fit (left panel).  
The derived mean velocity is $15.0 \pm 0.3$ km s$^{-1}$, similar to values of $\Vp$ found for AR 10961.  
Variations of the velocities around the fitted values are comparable to those in AR 11589, with $\st= 3.7$~km~s$^{-1}$ and $<\so > = 1.6$~km~s$^{-1}$ while the values relative to the mean velocity are a factor two higher  than in the two previous ARs. Indeed, large velocities, deviating significantly from the fit, are observed when the AR is at approximately 0.1\,--\,0.3 in $X$ after CMP (indicated by the arrows in Figure \ref{results_fig}g,h).  
This position corresponds to the two times when filaments erupted from the AR.  
Both eruptions appear to occur in the vicinity of the southern (FP) upflow region close to the EIS raster times.  
The effect of the filament eruptions is clearly indicated by the significant deviation of the velocity data points from the steady-flow model fit in Figure \ref{results_fig}g and in the increase in variation around the fitted values up to 15 km s$^{-1}$ in Figure~\ref{results_fig}h.  
The deviation from the model fit is evidence of intrinsic flow evolution in AR 10953. Notwithstanding the filament eruptions, upflows are very stable during the AR limb-to-limb transit. 

\subsection{General Results}
\label{general}
We have shown the model fits for three examples in order to demonstrate the various conditions under which upflows persist in ARs.  
In fact, these results are general to the ARs in our study since all of them have stationary flows over the LPs and FPs which are observed by \emph{Hinode}/EIS for 8--9 days during disk transit.
Stationary flows are likely to persist for much longer as the flows are seen in ARs with minimum ages that exceed the transit time. 
This is the case for half of the ARs in our study (see Table \ref{info} for AR minimum ages).

Derived mean velocities are in the range [9,41] km s$^{-1}$ and, where both polarities are observed by \emph{Hinode}/EIS, the velocities are similar (see Table \ref{results_tab}).  
Overall, the model fits are quite robust with the fluctuation of velocities around the fitted values typically within $\pm 1/10$ of the mean velocity, except when significant activity is present.
Intrinsic stationary flow evolution due to activity such as flux emergence, flaring, and eruptions is indicated by deviations from the model fits, as is  shown with AR 10953 (Figure \ref{results_fig}g and \ref{results_fig}h). 
Such periods of activity are easily identified from the residual velocities to the fit and are removed from the data set. Then the fit is performed again to derive the velocity parameters and the standard deviations outside the activity periods (Table~\ref{results_tab}).

The derived inclination angles of upflows for all ARs are given in Table \ref{results_tab}.
AR upflows typically fan away from the AR core by 40$\degree$ to near vertical (4$\degree$) for the dispersed FPs.  
The spread of inclination angles is more extensive for LPs with flows angled from 29$\degree$ inclined towards to 28$\degree$ directed away from the AR. 
However, with the exception of AR 10926 and AR 10961 (one of the two LP upflows), values of $\delta$ are consistent with those of the FPs, \ie\ flows are tilted away from the AR.

\section{Line Widths}
\label{ntvel}

Stationary flow model analysis of the Fe {\sc xii} spectral line width [$W$] was performed by fitting the data with 
   \begin{equation} \label{equation4}
   W = \Wo + \Wpr \sqrt{1 - X^2 - Y^2} - \Wph \frac{X}{1 - Y^2}
   \end{equation} 
where $X$ is the east-west normalized AR position, $\Wo$ 
is a constant which includes all width contributions independent of position, $\Wpr$ takes into account the flow dispersion in radial and latitudinal directions, 
similar to $V_{r}^{\prime}$ in Equation~(\ref{equation2}),  and $\Wph$ does the same in the longitudinal direction \citep[see Section 4.4 in][]{demoulin13}.  
Line widths free of instrumental broadening are given as FWHM in~\angstrom . 

\begin{figure}   
   \centerline{\includegraphics[width=1.0\textwidth,clip=0]{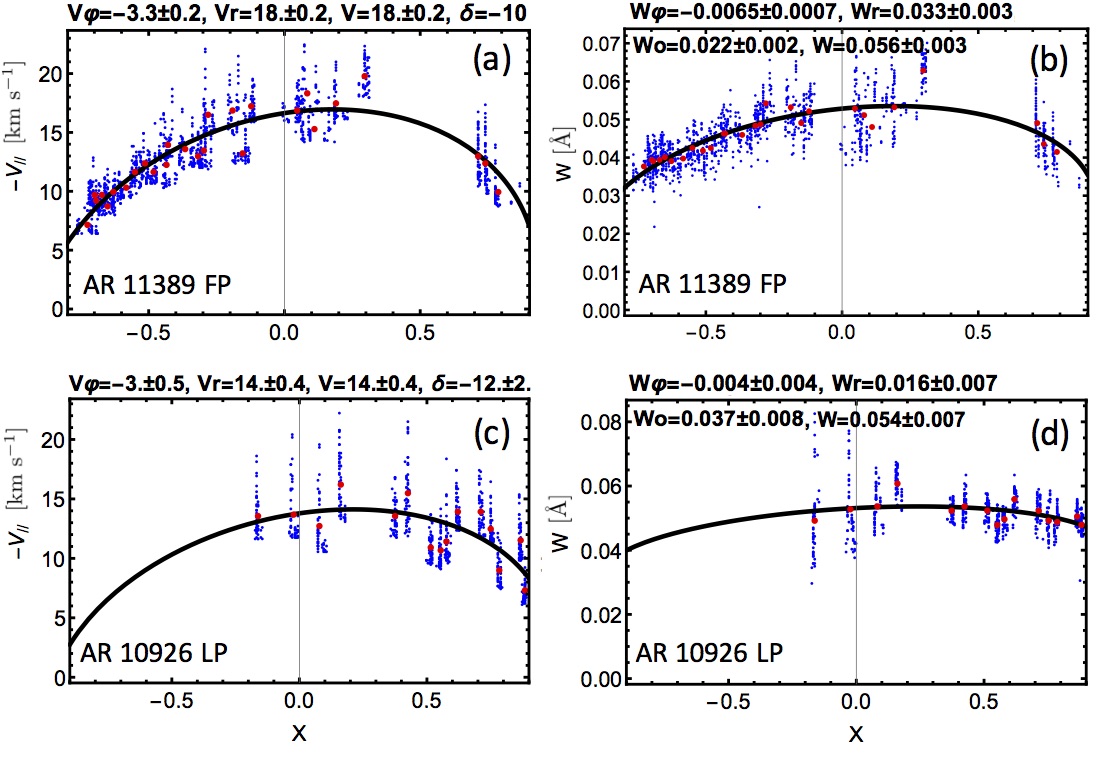}
              }
              \caption{
  Left panels: Steady-flow model results for AR 11389 FP and AR 10926 LP showing the dependence of the LOS upflow velocity [$\Vp$] on $X$ (east-west normalized AR position).
 Right panels:  Model results for line widths of AR 11389 FP and AR 10926 LP. See the caption to Figure \ref{results_fig} for the definition of the variables and the description of the plots.  AR 11389 FP shows strong rotational effects on both LOS velocity and line width, whereas AR 10926 LP presents weak effects for line width. 
 }
   \label{ntw}
   \end{figure}

Model fits for two extreme cases are shown in Figure \ref{ntw}.  
AR 11389 FP (Figure \ref{ntw}b and \ref{ntw}d) exhibits a relatively strong dependence of line width evolution on $X$, as is the case with LOS upflow velocity. 
Little activity was evident in this AR after its first few days on the disk \citep{baker15}.

Conversely, AR 10926 LP has a weak dependence of line width on $X$  (Figure \ref{ntw}d) compared to the dependence present in $\Vp$ (Figure \ref{ntw}c).
Significant and persistent flux emergence occurs when the AR is in the western hemisphere, thus introducing coronal perturbations which dominate the rotational effects in $W$.

The quantity $({\Wpr}^2+\Wph^2)^{1/2}/\Wo$ represents the relative effect of the rotation over $W_{0}$, which is the contribution to line width not affected by solar rotation.  
Columns 4 and 5 of Table \ref{results_tab} give values for the FP and the LP of all ARs where available.  
AR polarities with values approaching or exceeding unity have relatively strong rotational effects on line widths.
In such cases, these effects can be detected above $\Wo$, which includes thermal width (0.022~\angstrom\ for Fe {\sc xii}) and intrinsic fluctuations.
Occasionally, the variations of $W$ are falsely interpreted by the fit, \eg\ for AR 10942 FP which presents an upward-curved fit.  
For another example, AR 11575 (N) FP, the fits yield a non-physical result because $\Wo$ $<$ 0. 
This may occur when significant activity is present.

In general, the effect of rotation on $W$ appears to be weaker than on $\Vp$.
A case in point is AR 11564 which has strong apparent velocities of 30 km s$^{-1}$ for both polarities; however, the rotational effects are marginal for the FP, while they are strong for the LP.  Following the proposal of \cite{demoulin13}, the sensitivity of $W$ to rotational effects is expected to be due to the Doppler velocity differences within the same {\it Hinode}/EIS pixel.  
The distribution of the velocities increases the observed line width. For collimated flows, this effect has also a positional dependence due to the projection along the line of sight.
For many of the ARs in our study, the differential effect is weaker than the mean effect of the global line shift (due to the mean plasma velocity) and as a consequence may not be large enough to stand visible above $W_0$.  
This implies that rotational effects can be masked easily even without any activity.
The addition of activity increases $W_{0}$, and transforms $W_{r}^{\prime}$ and $\Wph$, masking further the effect of solar rotation. 
Finally, as is the case with $V$, we find no relation between the values of $W$ and the age of the ARs (\emph{c.f.} Table \ref{info} to Table \ref{results_tab}).

\section{Comparison of Deduced Inclinations and LFFF Extrapolation of AR 10926}
\label{extrap}

\begin{figure}   
   \centerline{\includegraphics[width=1.0\textwidth,clip=0]{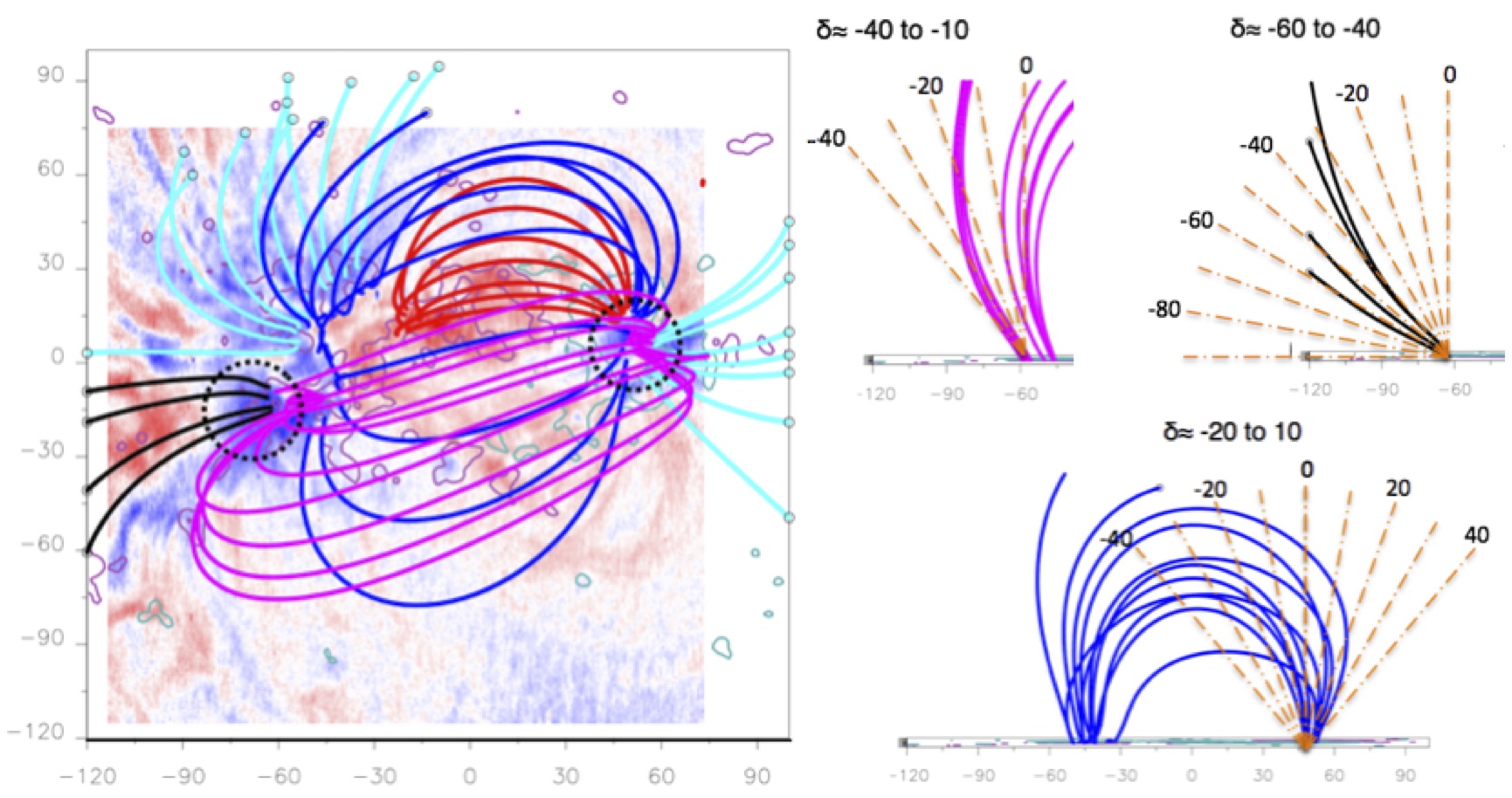}
              }
              \caption{Left panel:  Linear force-free field extrapolation for AR 10926 overlaid on {\it Hinode}/EIS Doppler velocity map at 16:28 UT on 1 December 2006.  Dashed circles indicate regions of strong upflows for FP and LP.  Right panels:  Inclinations of field lines rooted at locations of strong upflows for FP (top) and LP (bottom).}
   \label{extrap_fig}
   \end{figure}

We compute the magnetic field topology of AR 10926 during its disk transit on 30 November 2006 at 16:25 UT, 1 December at 16:28 UT, and 3 December at 15:32 UT.
The LOS magnetic field is extrapolated to the corona using a constant $\alpha$ linear force-free field (LFFF) configuration where $\myv{J} \times \myv{B} = 0$ and $\nabla \times \myv{B} = \alpha \myv{B}$ \citep{demoulin97}.  
Figure \ref{extrap_fig}, left panel, shows the LFFF model results for 1 December when the AR was close to CMP. Similar results are found at other times.
As in \cite{demoulin13}, the inclination deduced from the stationary flow model is compared with that of the LFFF extrapolation.
Field lines rooted in the strong upflow region of the FP form two groups, one of which connects to the LP within the AR and the other group connects outside of the computational box (pink and black field lines, respectively, in Figure \ref{extrap_fig}).
The ranges of inclination angles are [-40$\degree$, -10$\degree$] for the pink and [-60$\degree$, -40$\degree$] for the black field lines.
Unusually, compared to other ARs, inclination angles of the LP are within the range [-20$\degree$, 10$\degree$] with a number of field lines inclined to the east, \ie\ towards the center of the AR.

 The deduced inclinations [$\delta$] from EIS velocities reflect the asymmetry of the opposite polarities in the extrapolation and coronal observations with inclinations for FP of $\delta = -40 \pm 1\degree$ and for LP of $\delta = -12\pm 2\degree$, \ie\ both inclined toward the east.
Qualitatively, the steady-flow model and extrapolation  inclination angles are broadly comparable, however, there is a large range of inclinations in the extrapolations.

Sources of discrepancies between EIS upflow and magnetic extrapolation deduced directions may be due to a number of factors:\\ 
\emph{i}) We derive the mean inclination of the strongest flows over many days compared to a single snapshot of a model representing all AR field line populations.\\
\emph{ii}) The photospheric footpoints of the flows are difficult to estimate so the correspondence with the computed QSLs and the selected field lines is approximate.\\
\emph{iii}) Activity such as the flux emergence observed in this AR can influence the derived parameters of the steady-flow model.\\
\emph{iv}) LFFF extrapolations are computed assuming constant $\alpha$, however, the AR field may have varying $\alpha$ values as it evolves during disk transit.   
This is the case for AR 10926 (extrapolations not shown here).

The above source of discrepancies between EIS upflow directions and computed field lines are present for all ARs, with only differences in the relative importance of the above factors.   Globally we can only expect a broad coherence between the derived inclinations from both data sources. 

  \begin{figure}   
   \centerline{\includegraphics[width=1.0\textwidth,clip=0]{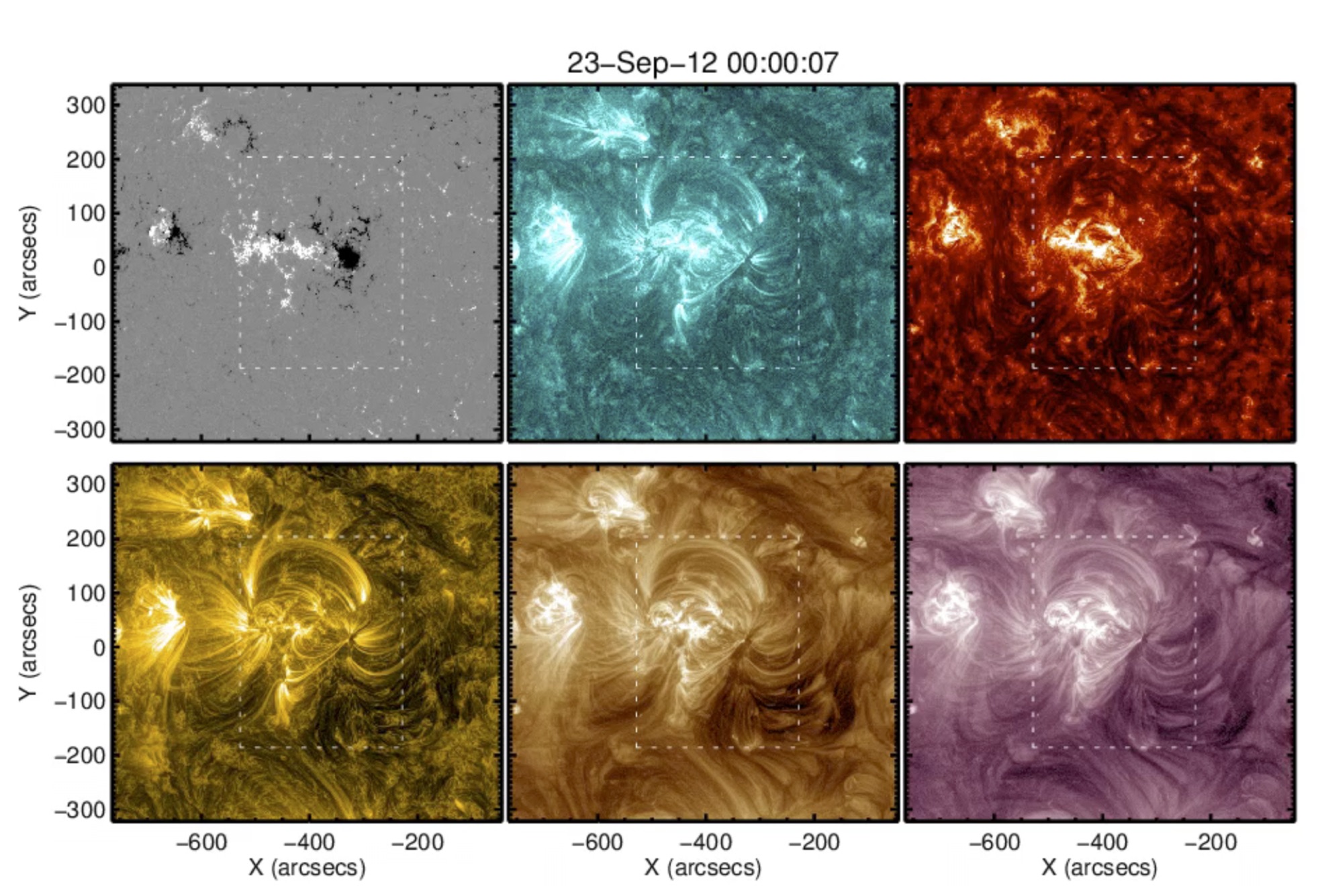}
              }

              \caption{Clockwise from upper left:  SDO/HMI magnetogram, SDO/AIA 131~\angstrom,  304~\angstrom, 211~\angstrom, 193~\angstrom, and 171~\angstrom~ images of AR 11575.  \emph{Hinode}/EIS FOV is indicated by dashed white boxes.  The image is extracted from Movie 2. }
   \label{pairs_aia}
   \end{figure}

\section{Upflow Pairs - AR 11575}
\label{pairs}
Three ARs in our sample have more than a single pair of upflow streams that are fitted with the stationary flow model.
AR 10953 flows can be divided into northern and southern streams.
AR 10961 has two upflow regions on the western side.
AR 11575 has four distinct flow pairs, one each in the northern, central, and southern areas within the AR, and one to the far south which is present only during the later EIS observations.   
Model results are provided in Table \ref{results_tab}.

  \begin{figure}   
   \centerline{\includegraphics[width=1.0\textwidth,clip=0]{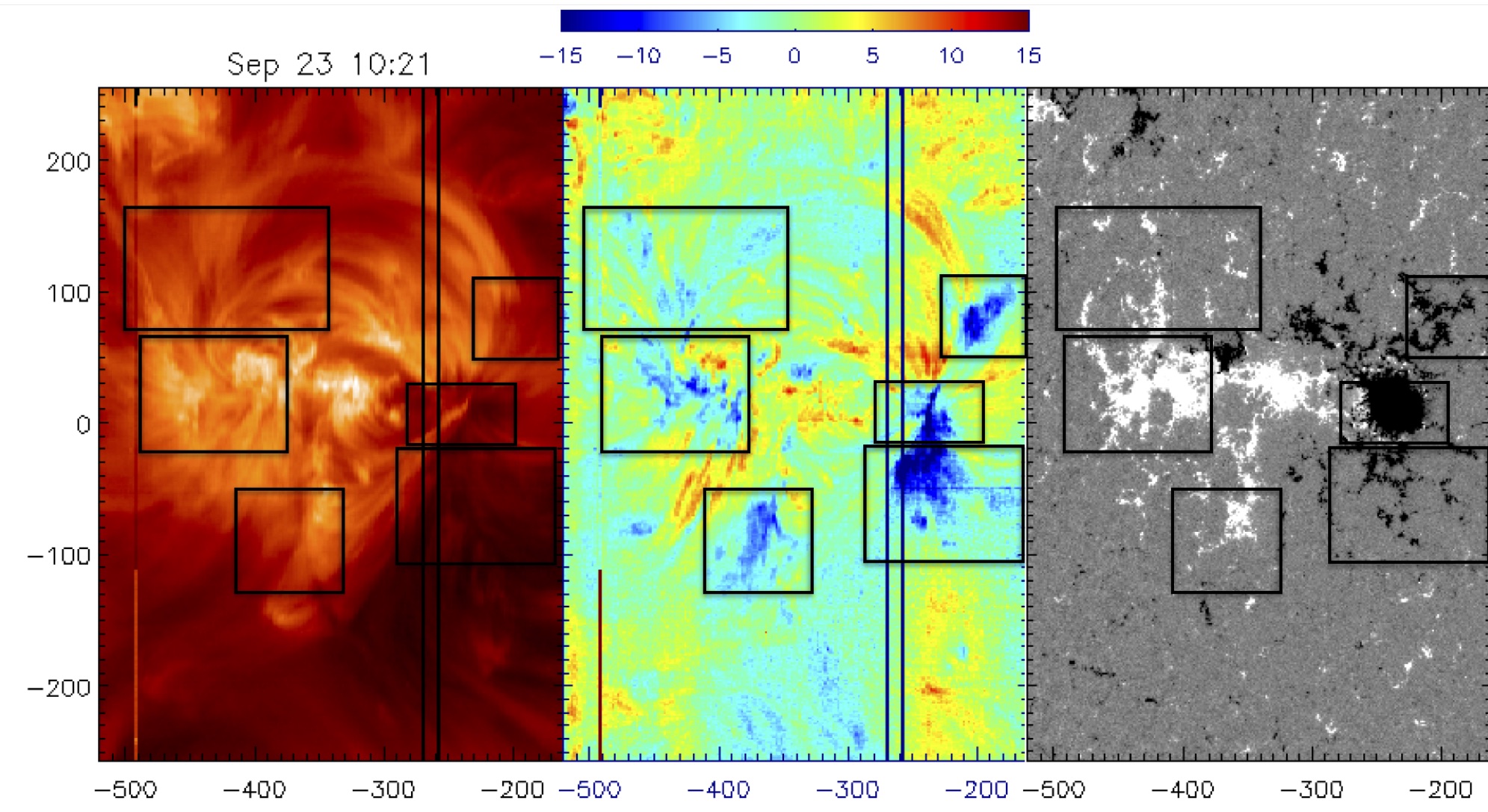}
              }
   \centerline{\includegraphics[width=1.0\textwidth,clip=0]{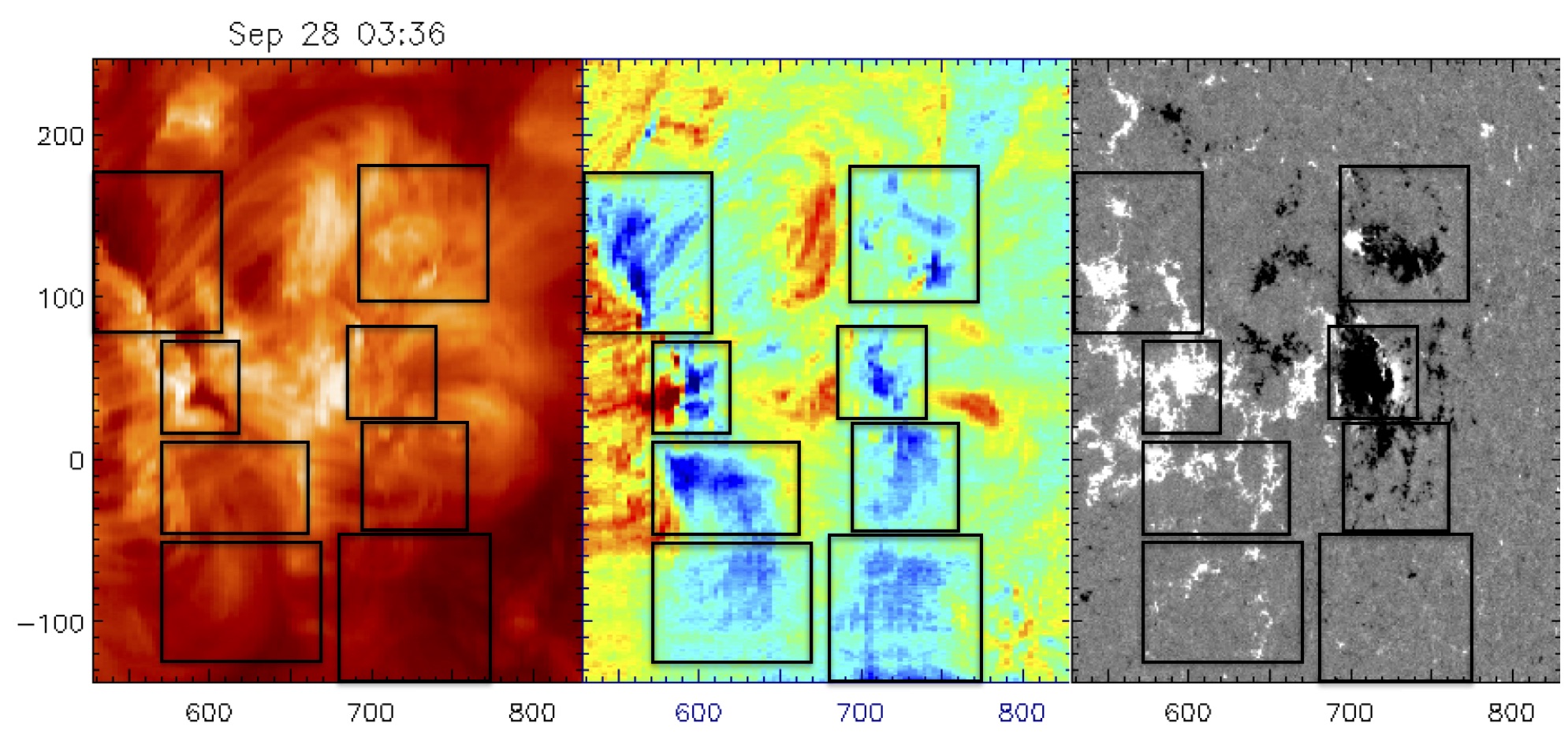}
              }
              \caption{EIS Fe {\sc xii} intensity (left), Doppler velocity maps (center), and SDO/HMI magnetogram (right) of AR 11575 on 23 (top) and 28 (bottom) September 2012.  Boxes indicate upflow pairs corresponding to different magnetic bipoles. Axes have units of arcsecs.  Color bar has units of km s$^{-1}$.}
   \label{pairs_fig}
   \end{figure}

Here we focus on AR 11575 which is featured in Figure \ref{pairs_aia} and Movie 2, both of which are composed of observations from the {\it Heliospheric Magnetic Imager} (HMI) and {\it Atmospheric Imaging Assembly} (AIA)  onboard the {\it Solar Dynamics Observatory} (SDO).
The AR is complex, formed by several bipoles which are evident in the HMI magnetograms.
In addition, loops rooted in the positive FP are connected to both the AR LP and to the negative polarity of the nearby AR 11577 located on the eastern side of AR 11575 (see Movie 2).
The strong connectivity between the ARs affects the eastern upflows, which differ from the typical flow regions observed in the isolated, simple bipolar ARs in our sample.
Parts of the upflows are located at the border of the extended FP rather than directly over the FP.

Individual flow pairs are identified in the Doppler velocity maps and are indicated by boxes in Figure \ref{pairs_fig} showing {\it Hinode}/EIS Fe {\sc xii} intensity and Doppler velocity maps with corresponding SDO/HMI magnetograms for 23 and 28 September.
On 23 September, there are three flow pairs and on 28 September, additional upflow streams appear together to the far south of the AR.
Upflows occur over pairs of opposite polarity magnetic field patches.
To the north, upflows occur at the base of long, close to potential loops connecting regions of dispersed magnetic field.
Short, compact loops within the AR core connect the intact leading spot with the FP.  To the south, the upflows occur in sheared loops connecting to very weak dispersed field. 

We are able to fit the model to the three pairs observed on 23 September but there are too few data to fit the southernmost pair.
Derived model velocities are highly comparable within these upflows with [FP, LP] velocities as follows: 
$V$ = [12, 12] for the northern pair, 
$V$ = [12, 16] for the central pair, and 
$V$ = [11, 23] for the southern pair. 
The derived $V$ values for the northern and southern pairs are likely to be  under-evaluated because we do not have measurements of the components of $V$ in the north\,--\,south direction.
Moreover, the model inclinations may not be well estimated due to the relatively low number of rasters available on the eastern side before CMP.
The near vertical inclination angle of the FP upflow of the southern pair (see Table \ref{results_tab})  is likely to be the result of a combination of the connectivity with AR 11577 and the low data coverage where $X<0$.

Despite the complex configuration of AR 11575, its flow pairs exhibit strong apparent evolution comparable to that of AR 11589 and AR 10961 shown in Figure~\ref{results_fig}.  
This is also the case with the separate flows and flow pairs identified in AR 10953 and AR 10961.  
In fact, for six of the ten ARs where both polarities are observed, the mean velocities derived from the model fits for each associated LP and FP are remarkably similar (see Section \ref{general} and Table \ref{results_tab}).  

Our results indicate that the origin of the upflow pairs is related.  It is unlikely that consistently similar flow velocities occurring on both sides of the ARs would be maintained for time scales of days/weeks unless the stationary flows are driven by a large-scale, global mechanism rather than a local one acting at a single polarity.  AR upflows originating from reconnection along QSLs between over-pressure AR loops and nearby under-pressure loops is consistent with such a mechanism as QSLs are defined by the global properties of the magnetic configuration of the AR and its surroundings \citep{demoulin07}.  
Moreover, \cite{mandrini15} showed that the upflows and QSLs evolve in parallel, both temporarily and spatially.

\section{Summary and Conclusions}
\label{end}

The aim of this study is to use {\it Hinode}/EIS data to constrain the geometry and nature of large-scale upflows present on both sides of ARs.  
We use the simple stationary-flow model applied to AR 10978 in \cite{demoulin13} to deduce the mean velocity [$V$], mean line width [$W$], and inclination angle [$\delta$] for nine ARs with reasonable limb-to-limb EIS coverage.  
By separating and characterizing apparent from intrinsic upflow evolution we set constraints on the mechanisms which are able to drive these large scale, persistent upflows. 
Our main results are:
\begin{itemize}
  \item All ARs in our study have stationary flows.  As a consequence, the observed long-term evolution of the collimated, large-scale, and persistent flows is largely due to the evolution of the LOS projection during disk transit. 
  \item In most cases, derived velocities [$V$] for the following and leading polarities are similar.  This holds for asymmetric ARs (\eg\ AR 10926 and AR 10961) indicating that magnetic field strength is not driving the stationary flows, \ie\ there is no direct magnetic acceleration like with CMEs.
  \item In general, ARs with stronger flows, \ie\ higher mean line-of-sight velocities, show stronger rotational effects. 
  \item Intrinsic flow evolution due to activity, such as flux emergence, jets, flaring, and CMEs, appears as a deviation from the model fit (\eg\ AR 10953, Figure~\ref{results_fig}g and \ref{results_fig}h). However, since such phenomena have different time scales, ranging from few tens of minutes (flares) to days (emergence), they affect differently the observed velocity.  They also affect the estimated steady flow with different amplitudes, CMEs having the largest imprint.  However, because of the different time scales, they can be identified in the velocity data and removed to derive the upflow properties.
  \item Our study demonstrates there is no evidence of a relationship between flow evolution and AR age which is consistent with the results of \cite{zangrilli16} who found with SOHO/UVCS that upflows remain for the entire lifetime of an AR.
  \item The effect of solar rotation on line width [$W$] is clearly present but weaker compared with its effect on $V$ for the nine new ARs in our study.  
 This contrasts with the strong dependence of both $V$ and $W$ on AR position in the case of AR10978 \citep{doschek08,bryans10,demoulin13}.  
The line broadening dependence on AR position is likely to be due to a large dispersion of velocities along the main flow direction. 
This effect is weaker than the global line shift so that thermal width and activity can easily mask the rotational effects on $W$.
  \item For the set of studied ARs, in the following polarity $\delta$ = [0$\degree, 40\degree$] is tilted away from the AR core. The leading polarity inclinations have a greater spread with $\delta$ = [-33$\degree, 28\degree$] such that the leading-polarity upflows tilt both towards and away from the center of the AR.
  \item  Deduced inclination angles [$\delta$] for both polarities of AR 10926 are broadly comparable to those obtained from a LFFF extrapolation of the AR.  The $\delta$ results of the stationary-flow model reflect the asymmetric topology of the AR evident in the extrapolation and coronal observations.   
  \item Independent flow pairs identified in three more complex ARs display apparent evolution, similar to the simple, isolated bipolar ARs in the sample.
The upflows appear in pairs with similar velocities in the following and leading AR polarities which hints that the same process is taking place at both polarities of the same pair.
\end{itemize}

Blue-shifted upflows are a common feature of ARs observed by {\it Hinode}/EIS throughout its ten-year mission.  
Our results imply that the stationary component of upflows occurring on either side of ARs are in fact related.  Moreover, stationary flows occur in pairs whether the ARs are isolated bipoles common at solar minimum or more complex, interconnected multipolar regions observed during solar maximum.   
This result constrains the possible upflow driving mechanisms.
Mechanisms acting locally at one polarity, \eg\ waves, nanoflares, or jets, are unlikely to produce globally stationary upflows with the same characteristics unless they are synchronized by another mechanism such as magnetic reconnection. Indeed our results are in agreement with a model where reconnection occurs along quasi-separatrix layers (QSLs) between over-pressure AR loops and neighboring under-pressure loops \citep{baker09,bradshaw11,delzanna11,lvdg12,demoulin13,mandrini15}.


\begin{acks}
The authors would like to thank Prof. Lidia van Driel-Gesztelyi for fruitful discussions in preparing this manuscript and Dr. David Long for making the movies.  We thank the anonymous referee for their constructive and informative comments.  {\it Hinode} is a Japanese mission developed and launched by ISAS/JAXA, collaborating with NAOJ as a domestic partner, NASA and STFC (UK) as international partners. Scientific operation of {\it Hinode} is by the {\it Hinode} science team organized at ISAS/JAXA. This team mainly consists of scientists from institutes in the partner countries. Support for the post-launch operation is provided by JAXA and NAOJ (Japan), STFC (UK), NASA, ESA, and NSC (Norway). DB is funded under STFC consolidated grant number ST/N000722/1. CHM acknowledges financial support from grants PICT 2012-0973 (ANPCyT), PIP 2012-01-403 (CONICET), and UBACyT 20020130100321. CHM is a member of the Carrera del Investigador Cient\'ifico (CONICET).
\end{acks}

\section*{Disclosure of Potential Conflicts of Interest}
The authors declare that they have no conflicts of interest.

\appendix
Table 4 gives a brief description of the EIS studies for the different active regions presented throughout the paper.

\begin{table*}[t]
\renewcommand{\arraystretch}{1.1}
  \caption{\emph{Hinode}/EIS study details for data used in this work.  Sparse raster is a scanning raster where the step size exceeds the slit width.
  }
\begin{tabular}{cccccc}
  \hline
Study No. & FOV  & Exposure Time & Slit & Total Raster Time &Comments\\
 &(arcsec)   & (seconds)  & (arcsec) & (hours)&\\
\hline
$3$ &
$256\times 512$&
$60$&
$1$&
$4.3$&
\\
$5$ &
$256\times 256$&
$30$&
$1$&
$2.1$&
\\
$7$ &
$512\times 256$&
$30$&
$1$&
$4.3$&
\\
$9$ &
$256\times 256$&
$10$&
$1$&
$0.7$&
\\
$10$ &
$256\times 256$&
$40$&
$1$&
$2.8$&
\\
$36$ &
$240\times 240$&
$5$&
$1$&
$0.3$&
\\
$37$ &
$240\times 240$&
$5$&
$2$&
$0.2$&
\\
$45$ &
$128\times 372$&
$60$&
$1$&
$2.1$&
\\
$46$ &
$256\times 256$&
$15$&
$1$&
$1.1$&
\\
$50$ &
$128\times 128$&
$25$&
$1$&
$0.9$&
\\
$84$ &
$120\times 256$&
$20$&
$1$&
$0.7$&
\\
$85$ &
$256\times 256$&
$10$&
$1$&
$0.7$&
\\
$95$ &
$257\times 408$&
$30$&
$1$&
$2.1$&
\\
$174$ &
$82\times 400$&
$25$&
$2$&
$0.3$&
\\
$176$ &
$82\times 400$&
$25$&
$2$&
$0.3$&
\\
$179$ &
$100\times 240$&
$25$&
$2$&
$0.3$&
\\
$180$ &
$180\times 512$&
$50$&
$2$&
$0.8$&
Sparse
\\
$261$ &
$360\times 512$&
$70$&
$2$&
$3.5$&
\\
$361$ &
$200\times 400$&
$10$&
$2$&
$0.3$&
\\
$381$ &
$195\times 280$&
$45$&
$2$&
$0.8$&
Sparse
\\
$437$ &
$240\times 512$&
$60$&
$1$&
$2.0$&
Sparse
\\
$454$ &
$303\times 384$&
$40$&
$2$&
$1.1$&
Sparse
\\
$471$ &
$360\times 512$&
$60$&
$1$&
$3.0$&
Sparse
\\
$480$ &
$480\times 512$&
$15$&
$1$&
$1.0$&
Sparse
\\

\hline
\end{tabular}
    \label{EIS_study}
\end{table*}

\bibliographystyle{spr-mp-sola}
\bibliography{3dflows}  



\end{article} 

\end{document}